%% file: paper.tex
\documentclass[letterpaper, 10pt, conference]{ieeeconf}
\IEEEoverridecommandlockouts
\usepackage{cite}
\usepackage{amsmath,amssymb,amsfonts}
\usepackage{ntdmath}
\usepackage{sequat}
\usepackage{graphicx}
\usepackage{textcomp}
\usepackage{tcolorbox}
\usepackage{cite}
\tcbset{colback=gray!5!white,colframe=gray!75!black,fonttitle=\bfseries}
\usepackage{subcaption}
\usepackage{tabularx}
\usepackage{booktabs}
\usepackage{multirow}
\usepackage{multicol}
\usepackage{wrapfig}
\usepackage{caption}
\usepackage{cuted}
\usepackage[
colorlinks,
bookmarksopen,
bookmarksnumbered,
citecolor=gray,
urlcolor=gray,
linkcolor=gray]{hyperref}
\usepackage{tikz}
\usepackage{minesvisual}
\usepackage{petrinet}
\usepackage{mathtools}
\usepackage[algoruled,vlined,linesnumbered]{algorithm2e}
\usepackage{stmaryrd}
\usepackage{arydshln}
\usepackage{xcolor}
\usepackage{colortbl}
\usepackage{dyatex}
\usepackage{mathtools}
\def\BibTeX{{\rm B\kern-.05em{\sc i\kern-.025em b}\kern-.08em
    T\kern-.1667em\lower.7ex\hbox{E}\kern-.125emX}}
\begin{document}

\title{\LARGE\bf Structural Analysis of Hybrid-Trace Nets: An Optimization Approach}

\author{Nhat Le \and Claire N. Bonial \and John G. Rogers \and Neil T. Dantam}
% \IEEEauthorblockA{\textit{Department of Computer Science} \\
% \textit{Colorado School of Mines}\\
% nguyencongnhat\_le@mines.edu}

\maketitle

\newcommand{\mpn}{HTrN\xspace}

\newcommand{\place}{p}
\newcommand{\placevar}[1]{\place_{#1}}
\newcommand{\trans}[1]{t_{#1}}
\newcommand{\transvar}[1]{\tau_{#1}}
\newcommand{\marking}[0]{\mathbf{m}}
\newcommand{\transvector}[0]{\boldsymbol{\tau}}
\newcommand{\incidence}[0]{\mathbf{C}}
\newcommand{\slackpmtx}{\mathbf{S}^+}
\newcommand{\slacknmtx}{\mathbf{S}^-}
\newcommand{\colspace}[1]{\mathbf{C}\left(#1\right)}
\newcommand{\nullspace}[1]{\mathbf{N}\left(#1\right)}
\newcommand{\reachablespace}{\mathbf{P}}
\newcommand{\mtxa}{\mathbf{A}}

\newcommand{\markingstep}[1]{\discstep{\mathbf{m}}{#1}}
\newcommand{\markingstepvar}[2]{\discstep{m_{#2}}{#1}}
\newcommand{\pset}{\mathcal{P}}
\newcommand{\bpset}{\mathcal{P}_b}
\newcommand{\rpset}{\mathcal{P}_c}
\newcommand{\ipset}{\mathcal{P}_d}
\newcommand{\tset}{\mathcal{T}}
\newcommand{\aset}{\mathcal{A}}
\newcommand{\baset}{\mathcal{A}_b}
\newcommand{\raset}{\mathcal{A}_c}
\newcommand{\iaset}{\mathcal{A}_d}
\newcommand{\wbfunc}{\mathcal{W}_b}
\newcommand{\wrfunc}{\mathcal{W}_c}
\newcommand{\wifunc}{\mathcal{W}_d}
\newcommand{\lb}{\ell}
\newcommand{\ub}{u}

\newcommand{\wfunc}{w}

\newcommand{\numvector}{\prescript{n}{}{\mathcal{M}}}
\newcommand{\numidn}{\incidence}
\newcommand{\slackp}[1]{s^+_{#1}}
\newcommand{\slackn}[1]{s^-_{#1}}
\newcommand{\slackpv}{\mathbf{s}^+}
\newcommand{\slacknv}{\mathbf{s}^-}
\newcommand{\relaxedtransvec}{\boldsymbol{\Tilde{\tau}}}

\newcommand{\temdomain}{\mathcal{X}}
\newcommand{\temrange}{\mathcal{Y}}

\newcommand{\pre}[1]{\mathrm{pre}_+(#1)}
\newcommand{\eff}[1]{\mathrm{eff}_+(#1)}
\newcommand{\trap}[1]{T(#1)}

\newcommand{\placeweight}[0]{\mathbf{w}}
\newcommand{\relaxedincidencemtx}{\mathbf{D}}
\newcommand{\rinmatxcell}[2]{\relaxedincidencemtx_{\left[{#1},{#2}\right]}}

\newcommand{\row}[1]{\mathbf{D}^R_{#1}}
\newcommand{\frp}{\mathcal{R'}}

\newcommand{\rowf}[2]{(#1)^R_{#2}}

\newcommand{\transboundcoef}[1]{\mu_{#1}}
\newcommand{\placeuppercoef}[1]{\lambda_{#1}^U}
\newcommand{\placelowercoef}[1]{\lambda_{#1}^L}
\newcommand{\trapcoef}[1]{\lambda_{#1}^T}
\newcommand{\trapsum}[3]{\raisebox{0.3ex}{$\textstyle -\!\!\!\sum\limits_{#2\in\mathcal{S}_{#1}}$}\!\!\rinmatxcell{#2}{#3}}
\newcommand{\sumterm}[3]{\raisebox{0.3ex}{$\textstyle\!\sum\limits_{#1}^{#2}$}#3}
\newcommand{\sumtermd}[5]{\raisebox{0.3ex}{$\textstyle\!\sum\limits_{#1}^{#2}$}\rinmatxcell{#3}{#4}#5}
\newcommand{\sumsumterm}[3]{\raisebox{0.3ex}{$\textstyle \!\!\sum\limits_{#1}^{#2}$}\;#3}
\newcommand{\trapmarkingsum}[3]{\raisebox{0.3ex}{$\textstyle\!\!\!\sum\limits_{#2\in\mathcal{S}_{#1}}$}\!\! \markingstepvar{#3}{#2}}
\newcommand{\trapmarkingsump}[3]{\raisebox{0.3ex}{$\textstyle\!\!\!\sum\limits_{#2\in\mathcal{S}_{#1}}$}\!\! \markingstepvar{#3}{#2}}

\newcommand{\mregion}{\mathcal{R}}
\newcommand{\uv}{\mathbf{u}}
\newcommand{\lv}{\boldsymbol{\ell}}

\newcommand{\gl}[1]{\mathcal{L}_{G}(#1)}
\newcommand{\ml}[1]{\mathcal{L}_{M}(#1)}

\newcommand{\transvarstep}[2]{\discstep{\transvar{#1}}{#2}}

\newtheorem{definition}{Definition}
\newtheorem{lemma}{Lemma}
\newtheorem{corollary}{Corollary}

\newcommand{\genrel}[0]{\mathit{op}}

\newcommand{\decvar}[1]{\textcolor{blue}{#1}}

\newcommand{\fcoefv}{\boldsymbol{\lambda}}
\newcommand{\cmf}{\decvar{\discstep{\marking}{h}}}

\newcommand{\firingvector}{\mathbf{z}}
\newcommand{\cfiringvector}{\decvar{\mathbf{z}}}

\newcommand{\fcoef}[1]{\boldsymbol{\lambda}_{#1}}
\newcommand{\cplaceweight}{\decvar{\placeweight}}

\newcommand{\cfcoef}[1]{\decvar{\fcoef{#1}}}
\newcommand{\cfcoefp}[1]{\decvar{\fcoef{#1}'}}
\newcommand{\ctrap}[1]{\decvar{\trap{#1}}}
\newcommand{\best}[1]{\textbf{#1}}
\begin{abstract}
% Petri nets (PNs) are a successful and well-known mathematical and graphical representation for modeling dynamic and concurrent event systems. PNs, however, only support nonnegative unbounded decision variables. We introduce a new variation of Petri nets with extensions for real bounded variables and Boolean transitioning semantics to (1) compactly represent discrete event systems and (2) effectively perform reachability analysis. We first show the formalism of our nets and propose a constraint-based approach for invariant analysis. Finally, we demonstrate the modeling power of our nets and the effectiveness of our analysis in task planning.

% Our PN model provides a reachability approximation, useful for identifying linear invariants. In addition, we introduce a new variant of trap constraints which strengthen our analysis. Finally, we provide an enumeration algorithm for extracting linear invariants through .

% discrete event systems modeled as trace nets
% many control problems can be reduced to reachability
Fault and error detection is crucial for ensuring the correctness of safety-critical systems.
Safety verification often addresses reachability of bad system states,
where
exhaustive state inspection may be intractable due to the large state
spaces of complex systems.
We propose a reachability relaxation for hybrid trace nets, a useful mathematical model for analyzing concurrent event systems, to efficiently verify system safety.
We further use our relaxation to effectively identify invariants, expressions that hold in all reachable states, allowing state-space reduction for efficient verification.
Empirically, our approach proves unsafe states unreachable in $1.5\times$ more instances, generates $2.5\times$ more invariants with up to $2$ orders of magnitude faster generation times than the baselines in the tested problems.

\end{abstract}

\section{Introduction}
% Analysis is hard
% invariants
% -> verifying safety/reachability (forbidden markings are not reached)
% -> synthesize control
% structural analysis
% program verification
% testing reachability is undecidable
% behavioral specification
% optimization
% structural analysis (invariants)
% behavioral analysis (reachability through enumerative methods vs state-equation methods)
% unsafe states, reachabiltiy
% forbiddence state, deadlock avoidance
% diagnosis

% continuous/hybrid nets
% trace nets
% place capacity nets
% non-monotonic behaviors, logical primitives
% unified framework

% Paragraph 1: Why do we care about trace nets reachability?
%
% Paragraph 2: Why is trace net reachability and invariant determination a worthy challenge?
%
% Paragraph 3: Thesis statement and outline of contributions.

% trace nets + hybrid
% high-level nets \cite{genrich1981system,kurt2009coloured}
\newcommand{\mpns}{HTrNs\xspace} Reliable systems should be free from
errors such as deadlocks or forbidden states.
Detecting and preventing errors such as unsafe states
% in safety-critical systems
is a key task in formal verification and synthesis.
Petri nets (PNs) are a formalism for event-driven models offering
effective tools for reachability and safety analysis.  PN variations
are well-studied, including discrete \cite{peterson1977petri},
continuous \cite{david2005discrete}, and hybrid
nets\cite{david2005discrete}.  Trace nets \cite{badouel1995trace} are
discrete nets with Boolean states and are useful for modeling systems
when events may modify state independently of prior states (e.g.,
servicing a machine guarantees a proper working status regardless of
the machine's prior conditions).
% (e.g., servicing a machine guarantees a proper working status regardless of the machine's prior conditions).
Analysis for trace nets (TNs), however, is less studied, especially in
a hybrid setting: continuous, integer, and Boolean (\mpns).

% second paragraph on what reachability analysis is, what makes it hard, what do we want out it
Deciding \mpn reachability is computationally challenging
\cite{cheng1995complexity,czerwinski2022reachability}. \mpn{}s induce
exponentially large state spaces, posing limits in testing
reachability.
However, PN structure provides two useful tools to address the
computational challenge: linear relaxation and invariants
\cite{murata1989petri}.
PN linear relaxations efficiently detect potentially reachable, and
definitely unreachable
states~\cite{murata1989petri}. %through a set of linear equations and inequalities
Invariants, which are linear expressions that hold in all reachable
states, improve empirical runtime of reachability tests by, for
example, reducing the number of decision
variables~\cite{spranger1998combining,schmidt2003using}.  Although
linear relaxation is widely applied in PNs
\cite{murata1989petri,esparza2000verification,esparza2014smt}, such a
structure has not been characterized in TNs (\mpns).  In addition,
existing PN invariant works
\cite{sankaranarayanan2003petri,silva1996linear,ji2022affine} fail for
TNs as they miss TN reachable states, while the approach in \cite{colon2003linear} is
impractical due to the prohibitive cost. We address the computational
issue of \mpn reachability by providing a reachability relaxation and
effective invariant extraction.

% existing invariant approaches \cite{sankaranarayanan2003petri,silva1996linear} fail to exploit the inherent structure in TNs, yielding ineffective analysis, while the general invariant method for transition systems \cite{colon2003linear} is impractical due to the prohibitive cost.

% (PSPACE-complete for TN and ACKERMANN-complete \cite{czerwinski2022reachability} for \mpn)
% state early in the intro
% In addition, \mpn analysis in general poses computational limits due to the state-space explosion problem (ACKERMANN-complete \cite{czerwinski2022reachability}).

\dyathesis{We develop an efficient reachability relaxation and
relaxation-based invariant generation for hybrid trace nets.}  
A hybrid trace net is a combination of PN formalisms.
We describe existing formalisms (see \autoref{sec:background}) and define hybrid trace nets (see \autoref{sec:formalism}).
Our relaxation optimistically determines reachability (see
\autoref{ssec:markingregion}) and helps to find invariants (see
\autoref{ssec:inv-method}). We show an efficient polynomial-time
invariant procedure via the relaxation (see \autoref{inv-mutex}) and
comprehensive invariant extraction via mixed-integer programming (see
\autoref{inv-milp}).

\section{Related Works}
\subsection{PN Formalisms}
Various Petri net formalisms are used for modeling different event systems.
Petri nets \cite{peterson1977petri}, currently known as ordinary PNs, were first introduced as a mathematical model for discrete event systems.
Prior works extend PNs to allow for continuous \cite{david2005discrete}, timed \cite{wang2012timed}, or probabilistic transition steps \cite{wang2012timed}.
While PNs generally concern systems with an unbounded number of states, Boolean nets \cite{tredup2018towards} and PNs with capacity constraints \cite{wolfgang1985petri} ensure boundedness on state variables to model finite systems. High-level PNs \cite{genrich1981system,kurt2009coloured} incorporate non-numeric states and extended transition rules to compactly represent event systems.
Our work studies trace nets, a type of Boolean nets, in a hybrid setting with ordinary, continuous, and capacity formalisms.

% Trace nets \cite{badouel1995trace}, a type of Boolean nets, differ from the well-known ordinary $0-1$ PNs in that the boundedness of state variables are enforced by the transition rules instead of the structure of the nets.

% Petri nets have many extended formalisms discrete, Continuous, Boolean
% In ordinary PNs \cite{peterson1977petri}, the state transition is a discrete process.
% Continuous Petri nets permit progressive transitioning, resulting in continuous state variables \cite{david2005discrete}.
% Place/Transition nets with place capacity \cite{wolfgang1985petri}\\
% Boolean nets -> trace nets

% While PNs have the typical mathematical representation, High-level Petri Nets \cite{genrich1981system,kurt2009coloured} incorporate additional structure (both numeric and non-numeric) to increase the expressiveness while compactly representing the systems.

% Realistic discrete event systems naturally involve time. The notion of time is introduced in Timed Petri nets \cite{wang2012timed} where each transition has an associated firing delay (or delay interval) that enforces its firing time after being enabled. Stochastic Petri Nets \cite{wang2012timed} are a timed variation that applies distribution functions to sample transition firing times and probabilistically select a transition for subsequent firing.

\subsection{Reachability Analysis}
% There are two widely used approaches to testing PN reachability: state-equation-based and enumerative methods \cite{murata1989petri}.

% On the other hand, enumerative approaches \cite{pastor2001symbolic,ma2016basis,li2025calculation} are complete, but they pose computational challenges due to the prohibitive cost of state-space exploration (ACKERMANN-complete \cite{czerwinski2022reachability}).
Prior works have addressed reachability analysis in TNs.
Deciding TN reachability is PSPACE-complete \cite{cheng1995complexity}.
Existing approaches reduce empirical runtime and memory usage through  unfolding \cite{chatain2022pomset}, constraint satisfaction \cite{biere2021bounded}, and symbolic representation \cite{tovchigrechko2009efficient,thierry2015symbolic}.
Chatain et al. \cite{chatain2022pomset} and Tovchigrechko \cite{tovchigrechko2009efficient} construct a complete set of reachable states before verification, while Biere \cite{biere2021bounded} computes and verifies reachable states at a fixed depth incrementally.
ITS-tools \cite{thierry2015symbolic} performs a suite of incomplete reachability routines: random forward state-space search, net reduction, and classical PN linear relaxation \cite{thierry2022symbolic}, in parallel to its state space construction.
However,\cite{chatain2022pomset,tovchigrechko2009efficient,thierry2015symbolic,biere2021bounded} are computationally hard in the worst case, while incomplete procedures in \cite{thierry2015symbolic} fail on trace nets.
We present a sound and incomplete procedure for deciding TN reachability in polynomial time through a linear relaxation.

% Chatain et al. \cite{chatain2022pomset} constructs an intermediate PN representation for acyclic event sequences, capturing all reachable states.
% \cite{biere2021bounded} computes reachable states at a fixed depth incrementally.
% \cite{tovchigrechko2009efficient} grows the complete reachable set by manipulating a compact tree-based data structure branching on integer intervals of state variables.
% \cite{thierry2015symbolic} performs a suite of incomplete reachability routines: random forward state-space search, net reduction, and classical PN linear relaxation \cite{thierry2022symbolic}, in parallel to its state space construction.

% Discrete: BDD\cite{pastor2001symbolic}, Basis\cite{ma2016basis}, TGAL \cite{li2025calculation}
% continuous \cite{blondin2017logics}
TN analysis in a hybrid setting is less explored.
% Prior works primarily focus on ordinary PNs \cite{murata1989petri,ma2016basis,li2025calculation}.
Hybrid formalisms \cite{fraca2015hybrid,david2005discrete} jointly reason about ordinary and continuous PNs, agnostic of TNs. 
While \cite{tovchigrechko2009efficient,chatain2022pomset} analyze hybrid TNs with ordinary PNs, they are limited in finite systems.
We provide a comprehensive analysis of TNs in conjunction with ordinary, continuous, and capacity formalisms.
% Inexact methods \cite{murata1989petri,esparza2000verification,esparza2014smt} characterize an over-approximation of reachable states through the state equation, allowing quick verification of safety properties; however, the use of state equation may be limited due to its optimistic behavior in testing reachability.

% Although linear relaxation is widely applied in general PNs \cite{murata1989petri,esparza2000verification,esparza2014smt}, such a structure has not been characterized in TNs (\mpns).

\subsection{Invariant Analysis}
Invariants are expressions over state variables that hold in all reachable states.
There are generally two classes of invariants in PNs: generalized mutual exclusion constraints (GMECs) and P-invariants. The former are conditions that limit weighted sums over state variables, while the latter are weight vectors inducing constant weighted sums over state variables. GMECs and P-invariants are widely used as safety specifications for synthesizing controlled PNs from original PNs \cite{yamalidou1996feedback,giua1992generalized}. Our work solves a different problem: extracting GMECs and P-invariants that hold in original PNs.

Prior works find invariants through induction \cite{colon2003linear,sankaranarayanan2003petri,ji2022affine} and PN linear relaxation \cite{esparza2000verification,esparza2014smt}.
Induction approaches produce conditions for affine invariants to hold before and after each event occurrence in transition systems.
Colon et al. \cite{colon2003linear} directly handles transition rules in TNs, but the conditions for solving are nonlinear and hence impractical in large systems.
Following works address the nonlinearity in \cite{colon2003linear} for ordinary PNs \cite{sankaranarayanan2003petri} and by extracting a selective set of invariants through matrix-algebra \cite{ji2022affine}.
However, approaches in \cite{ji2022affine,sankaranarayanan2003petri} cannot directly handle all transition rules in TNs.
Moreover, the constraint formulation for solving in \cite{ji2022affine,sankaranarayanan2003petri} is rigid and unable to incorporate information inherent in TNs (e.g., the upper bound on Boolean state variables) for an effective analysis.
In contrast, relaxation-based invariant analysis presents advantages as it freely incorporates additional information. However, it requires an over-approximation of reachability and poses limitation due to ignorance of event ordering. Esparza et al. \cite{esparza2000verification,esparza2014smt} address the latter challenge for ordinary PNs with traps, sets of state variables with positive sums.
We characterize a linear relaxation and extend the notion of traps for TNs to effectively find invariants.

\newcommand{\npre}[1]{\mathrm{pre_-}\left(#1\right)}
\newcommand{\neff}[1]{\mathrm{eff_-}\left(#1\right)}
\newcommand{\proposition}[1]{p_{#1}}
\newcommand{\propositionset}{\mathcal{P}}

\section{Background} \label{sec:background}
\subsection{Petri Nets}
An ordinary Petri net (PN) \cite{peterson1977petri} is a bipartite
graph of \emph{places} and \emph{transitions} connected by weighted
\emph{arcs}.
% A token is a single unit.
Each place contains a non-negative number of \emph{tokens}. %  Each
% non-negative weighted
Arcs describe flows representing when transitions can \emph{fire} to
change the token count in a place.  A transition is \emph{enabled}
when its input places have a number of tokens at least the arc weight.
An enabled transition can fire, withdrawing from its input places and
depositing to its output places numbers of tokens equal to the arc
weights. A \emph{marking} is a vector of token counts in all places. A
marking is reachable if there is a sequence of firings to produce the
marking from the initial marking.

A continuous Petri net (CTN) \cite{david2005discrete} is a Petri net
with continuous token flows. Places in CTNs have real/rational amounts
of tokens. A transition is enabled when each input place has a
positive token count. Transitions may fire by a continuous quantity
per step, distributing continuous flows of tokens across places.

% A CTN is a relaxation of a PN as the CTN discards the integrality of token counts and transition firings in the original PN.

A trace net (TN) \cite{badouel1995trace} is a Petri net with Boolean
places and flow relations \FuncSty{input}, \FuncSty{output},
\FuncSty{set}, \FuncSty{reset}, \FuncSty{used}, \FuncSty{free}, and
\FuncSty{nop} (see \autoref{table:boolean}). Relation \FuncSty{input}
(\textit{resp.}  \FuncSty{output}) consumes (\textit{resp.}  produces)
tokens.  Relations \FuncSty{used} and \FuncSty{free} test a place
(filled and emptied, respectively) but do change token counts.
Unconditional relation \FuncSty{set} (\textit{resp.}  \FuncSty{reset})
fills (\textit{resp.} empties) a place regardless of its prior token
count. \FuncSty{nop} shows no relation between a place and a
transition.

\subsection{Mixed Integer Linear Programming (MILP)} \label{ssec:milp}
A MILP is an optimization problem with linear constraints over
continuous and discrete variables. Modern optimization solvers
\cite{gurobi} find solutions using the efficient branch-and-bound
(BnB) framework. BnB works by branching on a tree of optimistic
solutions, solved relaxing integrality of discrete variables to
continuous, while maintaining bounds (the best relaxed solution and
the best incumbent) on subsequent solutions to prune weak branches on
the tree.

\newcommand{\transfunc}{f}
\newcommand{\rtset}{\mathcal{T}_c}
\newcommand{\itset}{\mathcal{T}_d}
\newcommand{\btset}{\mathcal{T}_b}

\section{Hybrid Trace Nets} \label{sec:formalism}
% \input{images/pn-boolean}
\input{tables/boolean-table}

% summarize trace net instead
% We provide a formal definition of hybrid trace nets (\mpn), followed by the state transition dynamics.
We define hybrid trace nets (\mpn) with integer, real, and Boolean
places and combining the flow relations in ordinary, continuous, and
trace nets. \mpns also allow box constraints on markings for
boundedness and efficient optimization.

\subsection{\mpn Definition}
\begin{definition}
A \mpn is a tuple ⟨$\pset$, $\tset$, $\aset$, $\wfunc$, $\lb$, $\ub$, $\markingstep{0}$, $\transfunc$⟩, where

\begin{itemize}
    \item $\pset = \ipset \cup \rpset \cup  \bpset$ and $\ipset$,
      $\rpset$,  $\bpset$ are disjoint sets of integer, real, and
      Boolean places;
    \item $\tset = \itset \cup \rtset $ and $\itset,\rtset$ are
      disjoint sets of discrete and continuous transitions;
    \item $\aset = \iaset \cup \raset \cup  \baset$ and $\iaset$,
      $\raset$,  $\baset$ are disjoint sets of weighted arcs
      connecting transitions and  integer, real, and Boolean places,
      respectively;
    % \item $\aset \subseteq (\bpset \times \tset) \cup (\tset \times \bpset)$ is a set of boolean arcs connecting boolean places and transitions.
    % \item $\wrfunc: \raset \rightarrow \mathbb{R}$ maps a real arc to a real weight.
    % \item $\wifunc: \iaset \rightarrow \mathbb{Z}$ maps a integer arc to an integer weight.
    % \item $\wbfunc: \baset \rightarrow \{-1,1\}$ maps a Boolean arc to a unit weight.
    % \item $\wfunc: \aset \rightarrow \mathbb{R}$ maps a arc to a weight.
    \item $\wfunc: \temdomain \rightarrow \temrange$
      % is a piecewise function that
      maps arcs to weights, where
      $\left(\temdomain, \temrange\right)\in \left\{ \left(\iaset,
          \intset\right), \left(\raset, \realset\right),
        \left(\baset, \left\{-1, 1\right\}\right)\right\}$;
    \item
      $\lb: \pset \rightarrow \realset \cup \left\{-\infty\right\}$
      maps places to lower bounds;
    \item
      $\ub: \pset \rightarrow \realset \cup \left\{\infty\right\}$
      maps places to upper bounds;
    % \item $\guard{\tau}$ is a set of guard conditions of a transition $\tau$.
    \item $\markingstep{0}$ is a start marking;
    \item and $\transfunc$ is a dynamic function that describes how
      token counts in \mpn places change due to transitions.
\end{itemize}
\end{definition}

A \mpn distinguish between real, integer, and Boolean places. Real
(integer) places have real (integer) token counts, whereas Boolean
places have either $0$ or $1$ token. The distinction between place
types is due to different transition semantics (see
\autoref{ssec:dynamics}) and affects the complexity of
decision/optimization procedures.

% The separation between numeric and Boolean places is due to the difference in their transition semantics (see \autoref{ssec:dynamics}), while the division between continuous and discrete places results from their different impact on the complexity of decision/optimization procedures.

% Tokens and arc weights may have real values, allowing \mpn to directly model systems with negative states (e.g. back-orders, deficits, etc) and continuous states.

% Different place types associate with different arc weights. Arc weights with real (integer) places are real (integer) numbers, whereas those with Boolean places are strictly $-1$ or $1$, indicating conditions and updates on the places (\autoref{table:boolean}).

A place may have predetermined box constraints (lower bound and upper bound), indicating the valid range of its token count. Enforcing bounds is useful for modeling systems with bounded states (e.g., temperature and capacity).

\subsection{\mpn Dynamics} \label{ssec:dynamics}
The dynamic function $\transfunc$ maps a marking and a transition firing to a successive marking.

% \subsubsection{Transition Encodings}
Due to the distinct flow relations associating with numeric (real and integer) and Boolean places, we use different encodings for the dynamics of each place group: linear equations and propositional logic, respectively.

% For numeric places $\placevar{i}\in \rpset\cup \ipset$, we consider one-go transitioning that combines individual sequential firings of a transition.
Numeric places gain tokens from incoming transitions and lose tokens to outgoing transitions. The token count $m_{i}$ of a numeric place $\placevar{i}$ is updated as,
\begin{equation} \label{eq:tokenflow}
    \small
    \markingstepvar{k+1}{i} = \markingstepvar{k}{i}  + \! \sum_{j\in
      \mathrm{in}}\wfunc\left(\placevar{i},\trans{j}\right)\transvarstep{j}{k}
    - \! \sum_{j\in \mathrm{out}} \wfunc\left(\trans{j},
      \placevar{i}\right)\transvarstep{j}{k} \! ,
\end{equation}
where $\transvarstep{j}{k} \in \realsetnonneg$,
($\intsetnonneg$) is a firing
quantity of a continuous (discrete) transition $\trans{j} \in \rtset$
($\itset$) at step $k$.

% Let $\markingstepvar{k}{i} \in \mathbb{R}$ $\left(\markingstepvar{k}{i} \in \mathbb{Z}\right)$ be the token count in a real (integer) place $\placevar{i}\in \rpset$ ($\placevar{i}\in \ipset$) at step $k$.

% start with box constraints
Box constraints in places must hold before, during, and after a transition firing.
A transition $\trans{j}$ may fire with a firing quantity $\transvarstep{j}{k}$ at step $k$ under the following conditions,
\begin{equation}
\small
\begin{aligned}\label{ieq:num-enable-cond}
    \lb(\placevar{i}) &\leq \markingstepvar{k}{i}  && \hspace{-0.5em} \leq \ub(\placevar{i}) \\
     \lb(\placevar{i}) &\leq \markingstepvar{k}{i}  - \wfunc(\placevar{i}, \trans{j})\transvarstep{j}{k} && \hspace{-0.5em} \leq \ub(\placevar{i}) \\
    \lb(\placevar{i}) &\leq \markingstepvar{k}{i}  + \wfunc(\trans{j}, \placevar{i})\transvarstep{j}{k} && \hspace{-0.5em} \leq \ub(\placevar{i}) \\
    \lb(\placevar{i}) &\leq\markingstepvar{k}{i}  -
                        \wfunc(\placevar{i},
                        \trans{j})\transvarstep{j}{k} +
                        \wfunc(\trans{j},
                        \placevar{i})\transvarstep{j}{k} && \hspace{-0.5em} \leq
                                                            \ub(\placevar{i})
                                                            \; .
\end{aligned}
\end{equation}

% Box constraints in places impose conditions (enabling conditions) under which transitions can fire and limits (enabling degrees) to their firing quantities.
% A transition $\trans{j}$ is enabled with a firing quantity $\transvarstep{j}{k}$ at step $k$ if the token counts in $\trans{j}$'s neighbor places $\placevar{i}$ are within their box constraints before, during, and after the firing,
% For Boolean places, we use a logic-based encoding for flow relations.

% \autoref{table:boolean} shows the correspondence between TN flow relations and our graphical representation of Boolean relations.

% An neighbor transition of a Boolean place $\placevar{i}$ either has positive ($1$) weighted arcs or negative ($-1$) weighted arcs from (to) the place.

% In the case of positive arcs, we use the following encoding as the transition rules,

Boolean places update token counts according to TN flow relations in \autoref{table:boolean}. We encode relations \FuncSty{set}, \FuncSty{output}, \FuncSty{used} between a Boolean place $\placevar{i}$ and neighbor transitions,
\begin{equation} \label{p-encoding}
    \begin{aligned}
      &\forall j,\
      \discstep{\transvar{\FuncSty{set},j}}{k}  &&\implies \markingstepvar{k+1}{i} \\
      & \forall \ell,\
      \discstep{\transvar{\FuncSty{output},l}}{k}  &&\implies \lnot\markingstepvar{k}{i} \land \markingstepvar{k+1}{i} \\
       &\forall o,\
      \discstep{\transvar{\FuncSty{used},o}}{k} &&\implies \markingstepvar{k}{i} \land
                                                 \markingstepvar{k+1}{i},
    \end{aligned}
\end{equation}
where $\transvar{\FuncSty{set},j},\transvar{\FuncSty{output},\ell},\transvar{\FuncSty{used},o} \in \left\{0,1\right\}$ are binary firings.

For relations \FuncSty{reset}, \FuncSty{input}, \FuncSty{free}, we use a similar encoding as \eqref{p-encoding} with opposite polarity on token count variables.

Boolean places may participate in continuous \FuncSty{used} and \FuncSty{free} relations,
\begin{equation} \label{p-encoding-c}
    \begin{aligned}
     &\forall o,\
      \discstep{\transvar{\FuncSty{cused},o}}{k} > 0 &&\implies \markingstepvar{k}{i} \land \markingstepvar{k+1}{i},\\
       &\forall \ell,\
      \discstep{\transvar{\FuncSty{cfree},\ell}}{k} > 0 &&\implies
                                                           \neg\markingstepvar{k}{i}
                                                           \land
                                                           \neg\markingstepvar{k+1}{i}
                                                           \; ,
    \end{aligned}
\end{equation}
where $\transvar{\FuncSty{cused},o}, \transvar{\FuncSty{cfree},\ell} \in \left[0,1\right]$ is a continuous firing. Other TN flow relations cannot support continuity as they change the token counts of Boolean places and continuous relations must keep the token counts unchanged for integrality.

\section{\mpn Reachability Relaxation} \label{ssec:markingregion}

% \subsection{\mpn Reachability Relaxation}
We introduce a linear relaxation to efficiently approximate \mpn
reachable markings.  Linear relaxations are widely used in PN analysis
\cite{murata1989petri,esparza2000verification,silva1996linear,esparza2014smt}.
However, existing relaxations do not capture \FuncSty{set} and
\FuncSty{reset} flow relations in TNs as they cause varying,
non-linear change.  We extend PN relaxations to handle \FuncSty{set}
and \FuncSty{reset} relations.

% We show the use of the feasible region in invariant analysis in \autoref{sec:inv}.

% Deciding the reachability is computationally hard \cite{czerwinski2022reachability}.

% The PN marking equation provides an effective tool for finding invariants that address the performance issue in optimization approaches \cite{achterberg2020presolve}.

\newcommand{\crelaxedtransvec}{\decvar{\relaxedtransvec}}

We consider the linear relaxation of \eqref{eq:tokenflow} that
combines the token flows across multiple steps into a single step,
neglecting transition ordering. This relaxation offers an optimistic
approximation of reachable markings.
Let $\discstep{\marking}{h} \in \realset^n$ be the vector of place
token counts, $\incidence \in \realset^{n\times m}$ be an incidence
matrix where each element $\incidence_{ij}$ denotes the token change
of place $\placevar{i}$ when transition $\trans{j}$ fires,
$\discstep{\transvector}{k} \in \realset^m_{\geq 0}$ be the firing
vector at step $k$, and $\relaxedtransvec \in \realset^m_{\geq 0}$
be the sum of firing vectors across steps. The relaxation is,
\begin{equation}
  \cmf =
  \discstep{\marking}{0} +
  \sum_{k=0}^{h-1}
  \incidence\decvar{\discstep{\transvector}{k}}
  =
  \discstep{\marking}{0} + \incidence\crelaxedtransvec \; .
     \label{eq:relaxed-pn-dynamics-numeric}
\end{equation}
The relaxed markings in \eqref{eq:relaxed-pn-dynamics-numeric} are
subject to box constraints. Let $\uv, \lv \in \realset^n$ be vectors
of lower and upper bounds,
\begin{equation} \label{ieq:box}
    \lv \leq \cmf \leq \uv.
\end{equation}

% be more explicit about 6 being existed

The \FuncSty{set} (\FuncSty{reset}) transitions cannot be captured by
incidence matrix $\incidence$  alone.
These transitions may rebind the token count to $1$ ($0$) when the
place is already filled (emptied) rather than linearly increasing
(decreasing) as described by  $\incidence_{ij}$.
%
% behavior
% increments (decrements) the token count of a Boolean place while
% rebinding the token count to $1$ ($0$) when the place is already
% filled (emptied).
%
% Such varying change in token count cannot be
% captured by the incident matrix alone as each element
% $\incidence_{ij}$ has exactly one value.
%
If we interpreted a \FuncSty{set} transition as token-incrementing
and accumulate its firings, the right hand side of
\eqref{eq:relaxed-pn-dynamics-numeric} would be greater than $1$,
causing any following transition that decrements the place to miss a
reachable marking where the place is emptied (false).

\newcommand{\cslackpv}{\decvar{\slackpv}}
\newcommand{\cslacknv}{\decvar{\slacknv}}

We extend \eqref{eq:relaxed-pn-dynamics-numeric} to handle the
\FuncSty{set} and \FuncSty{reset} relations in \mpn to ensure all
reachable markings are included in the relaxation. We treat
\FuncSty{set} (\FuncSty{reset}) relations as token-incrementing
(decrementing) and add non-negative slack variables in
\eqref{eq:relaxed-pn-dynamics-numeric} to adjust the relaxed counts.
If there is a \FuncSty{set} (\FuncSty{reset}) relation between a
transition and a Boolean place $\placevar{i}$, we use a slack variable
$\slackn{i}$ ($\slackp{i}$) to decrement (increment) the relaxed
count. Let
$\slackpv, \slacknv\in \realsetnonneg^m$ be
vectors of slack variables $\slackp{i}$ and $\slackn{i}$. If a place
$\placevar{i}$ is numeric or Boolean with no \FuncSty{set} or
\FuncSty{reset} neighbor transitions, then
$\slackp{i} = \slackn{i} = 0$. Our relaxation for \mpn is,
\begin{equation}
    \cmf = \discstep{\marking}{0} + \incidence\crelaxedtransvec +
    \cslackpv - \cslacknv \;.
    \label{eq:relaxed-pn-dynamics}
\end{equation}
%
% give it the readers that 8 is the 7 with all variable is a vector instead. make it short.
We rewrite \eqref{eq:relaxed-pn-dynamics} as affine system,
% that combines decision variables
% on the right hand side of \eqref{eq:relaxed-pn-dynamics},
\begin{equation} \label{eq:tf-relaxed-dynamics}
  \small
  \begin{split}
    \cmf
    % &=
    %        \discstep{\marking}{0}  +
    %        \left[
    %      \MAT{C}
    %      \ \MAT{I}^{m\times{}m}
    %      \ \MAT{I}^{m\times{}m}
    %        \right]
    %        \left[
    %        \relaxedtransvec^T
    %        \ \left(\slackpv\right)^T
    %        \ \left(\slacknv\right)^T
    %        \right]^T \\
           &=
           \discstep{\marking}{0} + \mathbf{D}\cfiringvector,
  \end{split}
\end{equation}
where
\mbox{\scriptsize$\mathbf{D}$} =
$\left[\begin{smallmatrix} \MAT{C} &
    \MAT{I}_m & \MAT{I}_m
      \end{smallmatrix}
    \right]$
and
\mbox{\scriptsize$\firingvector$}
=
$\left[
\begin{smallmatrix}
  \relaxedtransvec^T
  & \left(\slackpv\right)^T
  & \left(\slacknv\right)^T
\end{smallmatrix}
\right]^T$.
% where $\slackpmtx,\slacknmtx \in \mathbf{R^{n\times n}}$  are diagonal matrices. If the place $\placevar{i}$ involves in a \FuncSty{reset} (\FuncSty{set}) relation, $\slackpmtx_{ii} = 1$ ($\slacknmtx_{ii} = -1$); otherwise, $\slackpmtx_{ii} = 0$ ($\slacknmtx_{ii} = 0$).

% and denote the set of potentially reachable states $\discstep{\marking}{h}$ and the set of feasible firing vectors $\firingvector$ in (\ref{eq:tf-relaxed-dynamics}) as $\mregion$ and $\frp$,

We apply this relaxation to find invariants in \autoref{ssec:inv-method}.

% \begin{equation}
% \small
% \mregion =
% \left\{
% \discstep{\marking}{h} \in \mathbb{R}^n
% \;\middle|\;
% \begin{aligned}
% &\discstep{\marking}{h} = \discstep{\marking}{0} + \mathbf{D}\firingvector, \ \firingvector \geq 0,\\
% &\lv \leq \discstep{\marking}{h} \leq \uv,
% \end{aligned}
% \right\}.
% \label{feasible-region}
% \end{equation}
\section{Invariant Synthesis} \label{ssec:inv-method}
% start with what people knows
% distinguish between what exists and what is new
% some may go to the background
% go from existing works, to what break for tns, to how we solve it
% justify the NP hard of invariant extraction (maximal mutex, existing references)
% what is the delta
We apply \mpn reachability relaxation to find invariants.
Invariants are expressions that hold in all reachable markings.
We consider invariants that are affine constraints over places,
\begin{equation}
   \marking^T\placeweight
   \ \genrel \
   d
   \; ,
   \label{eq:gen-inv}
\end{equation}
where $\placeweight$ is a coefficient vector, $d$ is a constant, and $\genrel$ may be $\leq$ or $=$.
The general form \eqref{eq:gen-inv} has several subclasses.
When $\genrel$ is $\leq$, \eqref{eq:gen-inv} represents a generalized
mutual exclusion constraint \cite{giua1992generalized}.
When $\genrel$ is $=$, the vector $\placeweight$ in \eqref{eq:gen-inv} is a P-invariant \cite{silva1996linear}.
If \eqref{eq:gen-inv} involves only Boolean places and $\placeweight$
is a $0\text{-}1$ vector, the inequality (equality) of form
\eqref{eq:gen-inv} is a $k$-cardinality invariant
\cite{morgado2014mscg} ($k$-hot invariant
\cite{debevere2023quadratic}), where $d$ is a natural number $k$.
% For naming convenience, we prefix the aforementioned types with $k$, denoting an integral value of $d$, and refer to the former and the latter as  and $k$-hot invariants, respectively.

% TODO: make explicit connection to cuts. Maybe connect to multi-valued logic.
% citation for cdcl and dpll
% citation fast downward
Invariants are important for the efficiency of PN analysis.
$k$-cardinality invariants are closely related to clique cuts \cite{atamturk2000conflict} and cover cuts \cite{wolsey2020integer} used in optimization solvers to efficiently find solutions.
$k$-hot invariants provide a compact encoding for PN state space implicitly during the search \cite{davis1962machine,marques2002grasp} and explicitly during the state-space construction\cite{spranger1998combining,schmidt2003using,helmert2006fast}.

Our invariant analysis presents advantages over prior works.
Induction approaches are either computationally expensive \cite{colon2003linear} or unable to directly account for TN flow relations while imposing strict constraint formulations for solving \cite{sankaranarayanan2003petri,ji2022affine}.
We take a relaxation-based method to freely incorporate linear constraints inherent in TNs, e.g, box constraints, to effectively find invariants.
Prior relaxation-based approaches \cite{esparza2000verification,esparza2014smt} are computationally hard and only applicable to ordinary PNs.
Our work (1) offers a polynomial-time procedure for finding a class of affine invariants and (2) extends the relaxation-based analysis in \cite{esparza2000verification,esparza2014smt} for TNs.

We show a polynomial-time construction of $1$-cardinality and $1$-hot invariants in \autoref{inv-mutex}. We then extract \emph{k}-cardinality and \emph{k}-hot invariants with MILP in \autoref{inv-milp}.

\subsection{Polynomial-time Extraction} \label{inv-mutex}
We find $1$-cardinality and $1$-hot invariants by identifying pairwise mutual exclusions (mutexes) among Boolean places and then combining them into $1$-cardinality invariants, which are possibly converted into $1$-hot invariants.

Two places are mutex when they cannot both be simultaneously true.
Places $\placevar{u}$ and $\placevar{v}$ must be mutex when the
following linear constraints are infeasible under \eqref{ieq:box} and
\eqref{eq:tf-relaxed-dynamics},
\begin{equation}
\markingstepvar{h}{u} = 1 \land \markingstepvar{h}{v} = 1 \; .
\end{equation}

% connect equations to (9)
We use the collected pairwise mutexes $(\lnot \placevar{u} \lor \lnot \placevar{v})$ to construct efficiently-solvable $1$-cardinality constraints.
Pairwise mutexes can yield a quadratic number of constraints that are expensive to solve~\cite{rintanen2012planning}.
Instead, we reduce the number of constraints by constructing $1$-cardinality invariants,
\begin{equation}
  \sum_{i\in \mathrm{group}} m_{i} \leq 1
  \label{eq:pn:placemutex}
  \; ,
\end{equation}

Constructing maximal $1$-cardinality constraints is PSPACE-complete~\cite{fisler2018fact}, so we take a greedy approach that progressively grows $1$-cardinality invariants. 
An invariant starts with a single Boolean place $\placevar{i}$.
We individually add $\placevar{i}$'s mutex places  to the invariant such that the added place is mutually exclusive with all current places in the invariant.
We add places in decreasing order of how many mutex places they shared with $\placevar{i}$.
The invariant stops growing when the mutex relation between the place being added and places in the invariant no longer holds.

% When we can add another variable that is mutually exclusive with all current places in the invariant.

From a $1$-cardinality invariant, we can in some cases identify a more precise and efficiently solvable $1$-hot invariant when some place in the group is initially true and every transition that disables a place in the group enables another place in the group. The resulting one-hot invariant is,
\begin{equation}
  \sum_{i\in \mathrm{group}} m_{i} = 1
  \; .
  \label{eq:pn:placemutex}
\end{equation}

% put in summary and link to B
As our approach for constructing maximal $1$-cardinality invariants is
incomplete (but polynomial in time), the resulting invariants may not
be maximal.
We provide a sound and complete method for finding maximal $k$-cardinality invariants with $k \geq 1$ from our relaxation in \autoref{inv-milp}.

% \begin{align*}
% &\markingstepvar{h}{u} = 1, &&\markingstepvar{h}{v} = 1, \\
% &\discstep{\marking}{h} = \discstep{\marking}{0} + \mathbf{D}\firingvector,
% &&\lv \leq \discstep{\marking}{h} \leq \uv,
% &&&\firingvector \geq 0.
% \end{align*}

\subsection{MILP Extraction} \label{inv-milp}
% relate 8 and 9: produce a set of linear systems that relate marking equation to the invariants
We find maximal $k$-cardinality and $k$-hot invariants from our relaxation through MILP.
%
% As the number of these invariants is exponential\cite{silva1996linear}, we extract invariants over maximal sets of places.
%
% Constructing maximal $k$-cardinality invariants from our relaxation (\ref{eq:tf-relaxed-dynamics}) where $k = 1$ is computationally hard \cite{fisler2018fact}, similarly for $k > 1$.
Enumerating invariants is computationally hard as there are exponentially many invariants \cite{silva1996linear,soliman2012invariants}.
We leverage computational advances in optimization solvers to extract invariants.
%
% Enumerating linear invariants is computationally hard \cite{sankaranarayanan2003petri,,colon2003linear}.

Our approach has three steps: construction, tightening, and
enumeration.
We construct a system of linear constraints that relate the coefficients $\placeweight$ and $d$ in the inequality invariant \eqref{eq:gen-inv} and our relaxation \eqref{eq:tf-relaxed-dynamics} in \autoref{inv-constrs}.
The use of relaxation, however, may miss many invariants due to its over-approximation of reachable markings.
We tighten our relaxation with traps, invariants denoting positive token sums over places \cite{murata1989petri}, from the structure of TNs \autoref{sec:traps}.
Finally, we enumerate maximal $k$-cardinality invariants by repeatedly solving and progressively updating the set of constraints over $\placeweight$ and $d$ in \autoref{sec:inv-extract}.
Simultaneously solving similar constraints with opposite signs on $\placeweight$ and $d$ produces $k$-hot invariants.

\subsubsection{Constructing Invariant Constraints} \label{inv-constrs}
We produce a set of constraints whose solutions represent linear invariants.
We rewrite our linear relaxation, \eqref{ieq:box} and \eqref{eq:tf-relaxed-dynamics}, strictly as a system of affine inequalities in \eqref{eq:relaxed-ieq} to apply Farkas' lemma.
We then show how the system \eqref{eq:relaxed-ieq} and the inequality \eqref{eq:gen-inv} that holds in \eqref{eq:relaxed-ieq} connect to the lemma.
Finally, we process the result provided by the lemma to obtain a set of constraints on the coefficients of \eqref{eq:gen-inv}.

An affine inequality \eqref{eq:gen-inv} is an invariant if it holds
for all feasible solutions to $\discstep{\marking}{h}$ in the following system of affine inequalities from \eqref{ieq:box} and \eqref{eq:tf-relaxed-dynamics},
% line up the
\begin{equation}
\small
\begin{array}{rcrcl}
   \cmf & - & \mathbf{D}\cfiringvector & \leq & \discstep{\marking}{0} \\
   -\cmf & + & \mathbf{D}\cfiringvector & \leq & -\discstep{\marking}{0} \\
        & - & \cfiringvector & \leq & \mathbf{0} \\
   \cmf  &  &  & \leq & \uv \\
   -\cmf  &  &  & \leq & -\lv \; .
\end{array}
\label{eq:relaxed-ieq}
\end{equation}

% An affine inequality (\ref{eq:gen-inv}) is an invariant if it holds for all $\marking$ that are reachable markings in our relaxation, i.e, feasible solutions to $\discstep{\marking}{h}$ in a below system of affine inequalities from (\ref{ieq:box}) and (\ref{eq:tf-relaxed-dynamics}),

We use the following corollary of Farkas'
lemma~\cite{schrijver1998theory} for affine systems to relate
invariant \eqref{eq:gen-inv} and system \eqref{eq:relaxed-ieq}.

\begin{corollary}[Corollary 7.1h in
  \cite{schrijver1998theory}] \label{corollary} Let the system of
  linear inequalities $\MAT{A}\VEC{x} \leq \VEC{b}$ have at least one
  solution, and suppose that the linear inequality
  $\VEC{c}\VEC{x} \leq d$ is an invariant holding for each $\VEC{x}$
  satisfying $\MAT{A}\VEC{x} \leq \VEC{b}$. Then there exists a
  non-negative vector $\boldsymbol{\lambda}$ such that
  $\VEC{c} = \VEC{A}^T\boldsymbol{\lambda}$ and
  $\VEC{b}^T\boldsymbol{\lambda} \leq d$.
\end{corollary}

The system  $\MAT{A}\VEC{x} \leq \VEC{b}$ represents our system  \eqref{eq:relaxed-ieq} and the invariant $\VEC{c}\VEC{x} \leq d$ denotes the inequality \eqref{eq:gen-inv} with the same $d$ on the right hand side.

We form matrix $\MAT{A}$ from the constants on the left hand side
of \eqref{eq:relaxed-ieq}, vector $\VEC{b}$ from the constants on
the right hand side of \eqref{eq:relaxed-ieq}, and vector
$\VEC{x}$ from the variables in \eqref{eq:relaxed-ieq},
\begin{equation}
\small
\begin{split}
     \mathbf{A}^T &= \begin{bmatrix}
        \mathbf{I} & -\mathbf{I}& \mathbf{0}  & \mathbf{I} & -\mathbf{I} \\
        -\mathbf{D}^T &  \mathbf{D}^T&- \mathbf{I} & \mathbf{0} & \mathbf{0}
    \end{bmatrix},\\
    \mathbf{b}^T &= \begin{bmatrix}
        \left(\discstep{\marking}{0}\right)^T & -\left(\discstep{\marking}{0}\right)^T & \mathbf{0} ^T & \uv^T & -\lv^T
    \end{bmatrix}, \\
    \mathbf{x}^T &= \begin{bmatrix}
        \left(\cmf\right)^T & \cfiringvector^T
    \end{bmatrix}.
\end{split}
\end{equation}

Vector $\mathbf{c}$ contains unknown coefficients in linear invariants over variables in $\markingstep{h}$ and $\firingvector$. As \eqref{eq:gen-inv} is an invariant over places, we let the coefficients associated with variables in $\firingvector$ be zeros,
\begin{equation}
\small
\mathbf{c}^T =
\begin{bmatrix}
    \cplaceweight^T & \mathbf{0}^T
\end{bmatrix}.
\end{equation}

We create a variable in the vector $\fcoef{}$ for each linear constraint in the system \eqref{eq:relaxed-ieq} and denote $\fcoef{}$ as a stack of five different non-negative vectors corresponding to five different groups of constraints in \eqref{eq:relaxed-ieq},
\begin{equation}  \label{eq:farkas-coef}
\small
    \fcoef{}^T = \begin{bmatrix}
    \cfcoef{E_1}^T & \cfcoef{E_2}^T& \cfcoef{N}^T & \cfcoef{U}^T & \cfcoef{L}^T
\end{bmatrix}\; ,
\end{equation}

From the corollary, the inequality \eqref{eq:gen-inv} is an invariant in the system \eqref{eq:relaxed-ieq} iff the following constraints hold,
\begin{equation} \label{ieq:farkas-i-cond}
    \mathbf{c} = \mathbf{A}^T\boldsymbol{\lambda} \ \text{and} \
    \mathbf{b}^T\boldsymbol{\lambda} \leq d \; ,
\end{equation}
where $\mathbf{c}, \mathbf{A}, \mathbf{b}, \boldsymbol{\lambda}, d$ are defined as above.

Through matrix transformation and variable elimination ($\fcoef{E_1}$,
$\fcoef{E_2}$, and $\fcoef{N}$) in \eqref{ieq:farkas-i-cond}, we
obtain the following linear constraints to solve for $\placeweight$ and
$d$,
\begin{equation} \label{ieq:farkas-cond}
\small
\begin{array}{r@{\,}c@{\,}r@{\,}c@{\,}r@{\,}c@{\,}l}
  \left(\markingstep{0}\right)^T\cplaceweight & +  &\left(\uv \! - \! \markingstep{0}\right)^T\cfcoef{U} & -  & \left(\lv\! - \! \markingstep{0}\right)^T\cfcoef{L} & \leq & \decvar{d}, \\
  \MAT{D}^T\cplaceweight & - & \MAT{D}^T\cfcoef{U} & + & \MAT{D}^T\cfcoef{L}  & \leq & 0.
\end{array}
\end{equation}

A feasible assignment to $\placeweight$, $d$, $\fcoef{L}$, and
$\fcoef{U}$ in \eqref{ieq:farkas-cond} yields an inequality invariant
of form \eqref{eq:gen-inv}.

The use of \eqref{ieq:farkas-cond} may miss many invariants because the linear relaxation over-approximates reachable markings. We use trap constraints to tighten the relaxation to find more invariants than using the relaxation alone in \autoref{sec:traps}.

\subsubsection{Tightenting Relaxtion with Traps} \label{sec:traps}
We introduce traps in TNs to tighten our relaxation for finding invariants.
A trap is a set of places that, once marked (i.e., some place in the
set has at least one token), remains marked for all successive
steps~\cite{murata1989petri}.
We consider traps that are marked in the initial marking, i.e., linear invariants \eqref{eq:gen-inv} of form,
\begin{equation}
    \sum_{\placevar{i} \in \mathrm{Trap}} m_{i} \geq 1 \; .
    \label{cstr:trap}
\end{equation}

We extract minimal traps in TNs and combine them with the system
\eqref{eq:relaxed-ieq} to produce a set of constraints relating
invariant coefficients in \eqref{eq:gen-inv} and ourrelaxation and
traps.

Our work extends the notion of traps proposed by Esparza et al. \cite{esparza2000verification,esparza2014smt} for TNs.
The use of relaxation \eqref{eq:tf-relaxed-dynamics} may miss many invariants due to its over-approximation of reachability.
Esparza et al. use traps to tighten the set of relaxed reachable markings in ordinary PNs.
However, TN structure and flow relations render the proposed derivation of traps  inapplicable.
Esparza et al. encode traps
by relying on the structure and flow relations in ordinary
PNs---incoming (outgoing) transitions produce (consume) tokens---which
does not hold in TNs, e.g., incoming (outgoing) transitions in a
\FuncSty{reset} (\FuncSty{output}) relation consume (produce) tokens.
We provide an encoding that accounts for the TN structure and flow relations.

We construct a set of constraints whose solutions produce traps.
Let $\trap{\placevar{i}}$ be binary variables denoting whether Boolean places $\placevar{i}$ belong to the same trap.
The following conditions encode traps in TNs through variables $\trap{\placevar{i}}$.

Firstly, since we consider traps that hold in all reachable markings, i.e, from the initial marking, at least one Boolean place in a trap has one token in the initial marking,
\begin{equation}
\begin{split}
    \bigvee_{\placevar{i} \in \bpset} \left(\ctrap{\placevar{i}}
  \wedge \markingstepvar{0}{i} = 1\right) \; .
    \label{exp:trap-init}
\end{split}
\end{equation}

Secondly, for every transition in a TN, if at least one place in a trap is required to be filled for firing, at least one place in the same trap must be filled after firing,
\begin{equation}
\begin{split}
     \bigvee_{\placevar{i} \in \pre{\tau} } \ctrap{\placevar{i}}
  \implies \bigvee_{\placevar{j} \in \eff{\tau}} \ctrap{\placevar{j}},
  \quad \forall \tau\in \tset \; ,
    \label{exp:trap}
\end{split}
\end{equation}
where $\mbox{\small{$\pre{\tau}$}} = \mbox{\small{$ \{\placevar{i} \in \bpset \ | \ \wfunc(\placevar{i}, \tau) = 1\}$}}$ and $\mbox{\small{$\eff{\tau}$}}  = \mbox{\small {$\{\placevar{i} \in \bpset \ | \ \left(\wfunc(\placevar{i}, \tau) = -1 \wedge \wfunc(\tau, \placevar{i}) = 0\right) \lor \left(\wfunc(\tau, \placevar{i}) = 1\right)\}$}}$.

% A set of places $\placevar{i}$ with $\trap{\placevar{i}} = 1$ satisfying constraints (\ref{exp:trap-init}) and (\ref{exp:trap}) forms a trap in a TN.

An assignment to variables $\trap{\placevar{i}}$ satisfying constraints \eqref{exp:trap-init} and \eqref{exp:trap} yields a trap where $\trap{\placevar{i}} = 1$ for every place in the trap and $\trap{\placevar{i}} = 0$ for any other place.

We find minimal traps to avoid the large number of traps\cite{der1996generating}. 
A trap is minimal if removing any place makes the remaining set not a trap. 
Enumerating minimal traps is computationally hard as there are exponentially many of them in the worst case\cite{nabli2012boolean}. 
We use optimization solvers to efficiently produce minimal traps. 
We repeatedly solve \eqref{exp:trap-init} and \eqref{exp:trap} while minimizing the sum over variables $\trap{\placevar{i}}$ and incrementally adding the following constraint for every newly solved trap to diversify solutions,
\begin{equation}
    \sum_{\placevar{i} \in \mathrm{Trap}} \ctrap{\placevar{i}} \leq
    |\mathrm{Trap}| - 1 \; .
    \label{cstr:min-trap}
\end{equation}

After finding all minimal traps, we apply the traps of form \eqref{cstr:trap} to tighten our relaxation for finding invariants.
We stack the minimal traps into a matrix,
\begin{equation}
    -\mathbf{N}\cmf \leq -\mathbf{1} \; ,
\end{equation}
where $\mathbf{N}$ is a $0\text{-}1$ matrix, and combine it into \eqref{eq:relaxed-ieq}.

We add a nonnegative vector $\fcoef{T}$, where each element associates with a minimal trap constraint, to $\fcoef{}$ in \eqref{eq:farkas-coef},
\begin{equation}
  \small
  \fcoef{}^T =
  \begin{bmatrix}
    \cfcoef{E_1}^T & \cfcoef{E_2}^T &  \cfcoef{N}^T & \cfcoef{U}^T & \cfcoef{L}^T & \cfcoef{T}^T
  \end{bmatrix}
\; .
\end{equation}

Following a similar derivation to \autoref{inv-constrs}, we obtain a
set of constraints to solve for $\placeweight$ and $d$,
\begin{equation} \label{ieq:farkas-cond-trap}
\small
\setlength{\arraycolsep}{1pt}
\hspace{-0.6em}
\begin{bmatrix}
     (\markingstep{0})^T & (\uv\! - \! \markingstep{0})^T & (\markingstep{0} \!\! - \lv)^T & (\mathbf{N}\markingstep{0}\!\! - \! \mathbf{1})^T \\
     \mathbf{D}^T & -\mathbf{D}^T & \mathbf{D}^T & \mathbf{D}^T\mathbf{N}^T
\end{bmatrix} \!\!\decvar{\mathbf{Y}} \! \leq \!
\begin{bmatrix}
    \decvar{d} \\
    0
\end{bmatrix} \!\!,  \\
\end{equation}
where \mbox{\scriptsize$\mathbf{Y}^T = \begin{bmatrix}
    \placeweight^T & \fcoef{U}^T & \fcoef{L}^T & \fcoef{T}^T
\end{bmatrix}$}.

We use constraints \eqref{ieq:farkas-cond-trap} to extract invariants in \autoref{sec:inv-extract}.

\subsubsection{Enumerating Invariants} \label{sec:inv-extract}
We apply the constraints \eqref{ieq:farkas-cond-trap} to enumerate maximal \emph{k}-cardinality and \emph{k}-hot invariants over Boolean places.
A $k$-cardinality invariant is maximal if flipping any zero element in
$\placeweight$ in \eqref{eq:gen-inv} makes the resulting expression
not a $k$-cardinality invariant.
We extract only maximal $k$-cardinality invariants to avoid redundancy.

We repeatedly solve a MILP to enumerate invariants.
%
% A solution to the MILP corresponds to either a maximal $k$-cardinality invariant or a $k$-hot invariant.
%
The MILP computes a maximal $k$-cardinality invariant and converts it to a $k$-hot invariant when possible.
For each found invariant, we incrementally add a constraint to the MILP for re-solving to ensure the uniqueness of subsequent solutions.
The process repeats until the MILP is infeasible.

The following MILP yields an invariant with a fixed $k$,
\begin{align}
\label{milp:obj}&\operatorname{lexmax} \bigl( \sum_{\placevar{i} \in \pset} \decvar{w_i},\; \decvar{y}\bigr) \\
s.t. \
\label{milp:coef-cstr} &\setlength{\arraycolsep}{3pt}
\begin{bmatrix}
    \cplaceweight^T & \cfcoef{U}^T & \cfcoef{L}^T & \cfcoef{T}^T
\end{bmatrix}\mathbf{M}^T \leq
\begin{bmatrix}
    \decvar{d} & 0
\end{bmatrix}\\
\label{milp:d} & k \leq \decvar{d} < k + 1, \\
\label{milp:ind-inv}& \decvar{y} \Rightarrow \setlength{\arraycolsep}{3pt}
\begin{bmatrix}
    -\cplaceweight^T & \cfcoefp{U}^T & \cfcoefp{L}^T & \cfcoefp{T}^T
\end{bmatrix}\mathbf{M}^T \leq
\begin{bmatrix}
    -\decvar{d'} & 0
\end{bmatrix}\\
\label{milp:ind-d}&\decvar{y} \Rightarrow k - 1 < \decvar{d'} \leq k,  \\
\label{milp:w-0} & \decvar{w_i} = 0, \quad \quad \quad \!\forall \placevar{i}\in \pset \backslash \bpset, \\
\label{milp:w-b} & \decvar{w_i} \in \{0,1\}, \quad \forall \placevar{i}\in \pset,\\
& \decvar{y} \in \{0,1\}, \\
& \decvar{d},\decvar{d'},\cfcoef{U}, \cfcoef{L}, \cfcoef{T}, \cfcoefp{U}, \cfcoefp{L}, \cfcoefp{T} \text{ are non-negative}.
\end{align}

The lexicographic objective \eqref{milp:obj} maximizes the total number of Boolean places appearing in a $k$-cardinality invariant and then maximizes a binary variable $y$, denoting whether the invariant becomes $k$-hot.
Constraints \eqref{milp:coef-cstr} relate invariant coefficients in \eqref{eq:gen-inv} to our relaxation and traps, same as \eqref{ieq:farkas-cond-trap}, where  $\mathbf{M}$ denotes the coefficient matrix on the left hand side of  \eqref{ieq:farkas-cond-trap}.
%
% A $k$-cardinality invariant is a special form of  \eqref{eq:gen-inv} where $d$ has an integer value $k$.
%
Constraint \eqref{milp:d} allows the term $d$ in \eqref{eq:gen-inv} to round down to $k$, potentially producing more invariants than fixing $d$ to $k$.
Indicator constraints \eqref{milp:ind-inv} and \eqref{milp:ind-d} enforce  implied constraints when $y$ is $1$.
Implied constraints in \eqref{milp:ind-inv} and \eqref{milp:ind-d} relate invariants of form \eqref{eq:gen-inv}, with sign-opposite coefficients to the coefficients yielded by \eqref{milp:coef-cstr} and \eqref{milp:d}, to our relaxation and traps.
When \eqref{milp:coef-cstr} and \eqref{milp:d} admit a $k$-cardinality invariant, the objective sets $y$ to $1$ when possible, activating implied constraints in  \eqref{milp:ind-inv} and \eqref{milp:ind-d} so that the invariant becomes \emph{k}-hot.
Implied constraints in \eqref{milp:ind-inv} follows a similar derivation of \eqref{ieq:farkas-cond-trap} in \autoref{sec:traps}.
We use $d'$ in the implied constraint \eqref{milp:ind-d} instead of $d$ as they have different relations to $k$.
Similar to \eqref{milp:d}, we allow $d'$ to round up to $k$ to find more invariants.
Finally, constraints (\ref{milp:w-0}) and (\ref{milp:w-b}) ensure only Boolean places ever appear in generated invariants.

To avoid redundancy and diversify solutions, we prevent places in subsequent invariants from forming a subset of places in a solved invariant. At least a place not in a MILP solution must be in the next solution. We incrementally add the below constraint to the MILP after each solve,

\begin{equation}
    \sum_{p_i\in \bpset \backslash \{p_i|w_i=1\}} \decvar{w_i} \geq 1
    \; .
\end{equation}

We evaluate our enumeration algorithm in \autoref{exp:inv}.

\newcommand{\pddlActions}{\ensuremath{A}}
\newcommand{\pddlaction}{\ensuremath{a}}
\newcommand{\pddlVars}{\ensuremath{V}}
\newcommand{\pddlvariable}{\ensuremath{v}}
\newcommand{\planstart}[0]{\ensuremath{I}}
\newcommand{\plangoal}[0]{\ensuremath{G}}

\section{Experiments and discussion} \label{sec:exp}
 We evaluate the effectiveness of our relaxation in \autoref{ssec:markingregion} in safety verification, i.e., proving unreachabilities of unsafe markings, and the invariants generated by our approaches in \autoref{ssec:inv-method}.
 We compare the use of our relaxation against the state-of-the-art model checker ITS-tools \cite{thierry2015symbolic} and our invariant generations against induction approaches \cite{colon2003linear}, implemented in StInG \cite{sting2004}, and \cite{sankaranarayanan2003petri}, implemented ourselves.
 StInG \cite{sting2004} allows affine invariant synthesis in trace nets and \cite{sankaranarayanan2003petri} is the leading method in PN literature.
 Our relaxation proves $1.5\times$ more unreachabilities than ITS-tools and our invariant generations produce up to $2.5\times$ more invariants and run up to $2$ order of magnitude faster than \cite{sting2004,sankaranarayanan2003petri} in the tested problems.

Our implementation extends TMKit~\cite{dantam2018tmkit} with Gurobi \cite{gurobi} to solve constraints.
%
% \eqref{ieq:box} and \eqref{eq:tf-relaxed-dynamics} in our relaxation and the MILP program for our invariant enumeration in \autoref{sec:inv-extract}.
%
The tested problems come from the past Model Checking Contest (MCC) \cite{mcc2026} and International Planning Competitions (IPC) \cite{ipc2016,ipc2000,ipc2014,ipc2023numeric}, or are generated ourselves.
We ran our benchmarks using one core of an AMD Ryzen 9 7950X3D CPU
under Debian 12.

% We select problems from MCC and IPC as they provide de-facto standard problems for evaluating the state-of-the-art model checking and planning systems \cite{amat2023behind,taitler20242023}.

MCC and IPC problems are computationally challenging and varied in system scenarios \cite{amat2023behind,taitler20242023}, forming a diverse benchmark suite.
MCC provides direct problems for safety verification in ordinary PNs.
IPC problems reason over Boolean- and numeric-valued state variables, which naturally map to hybrid trace nets, providing a good measure for evaluating our analysis.
An IPC problem consists of state variables, an initial state, a set of actions with preconditions and effects, and a set of goal states.
Places correspond to state variables; transitions correspond to actions; and arcs encode preconditions and effects and thus TN flow relations in \autoref{table:boolean}.
The initial state corresponds to the initial marking and goal states correspond to unsafe markings for verification.

\subsection{Safety Verification}

\begin{figure} [t]
    \centering
    \includegraphics[width=1\linewidth]{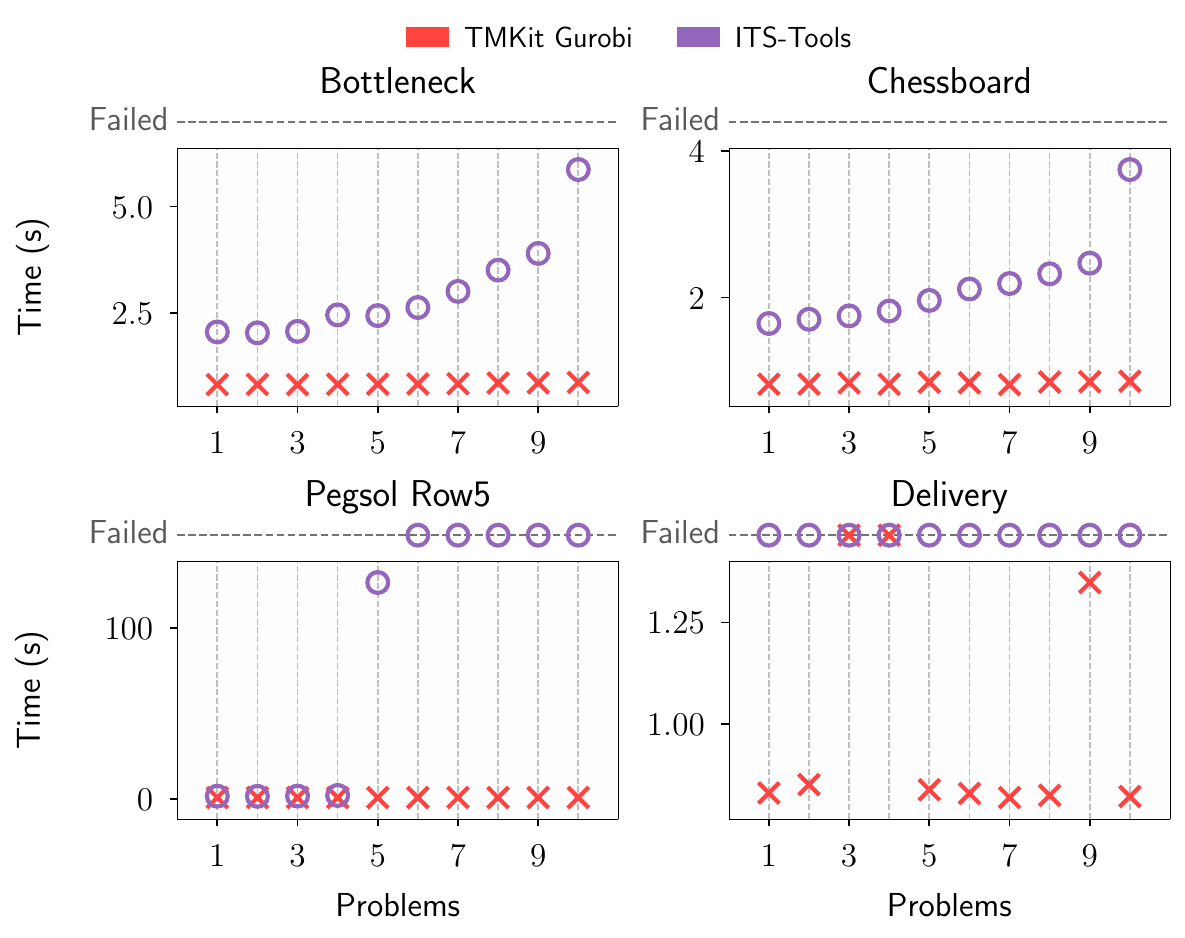}
    \caption{Safety verification times in four families of trace nets. Net size in each subplot increases from left to right. \emph{Failed} points represent $10$-minute timeout or $8$-GiB memory limit.}
    \label{img:inftime}
\end{figure}

\begin{table}[t]
\centering
\setlength{\tabcolsep}{3pt}
\begin{tabular}{l|c|c|c}
\hline
Families & Property Count & ITS & TMKit \\
\hline
Flexible Manufacturing System & 37 & \best{37} & \best{37} \\
Peterson & 34 & \best{34} & 29 \\
Robot Manipulation & 19 & \best{19} & 18 \\
Resource Allocation & 36 & \best{36} & 29 \\
\hline
Total             & 126 & \best{126} & 113 \\
\hline
\end{tabular}
\caption{Numbers of unreachabilities detected in four families of ordinary PNs. Excluded counts hit $10$-minute timeout or $8$-GiB memory limit.}
\label{table:mcc-false-ef}
\end{table}

% for safety verification are  properties in discrete event systems that
We evaluate our approach on a set of safety problems from MCC 2026 \cite{mcc2026}, IPC 2016 \cite{ipc2016}, and self-generated (Delivery \cite{ipc2023numeric}).
Event systems for safety verification are ordinary PNs in MCC 2026, trace nets in IPC 2016, and a mix of both formalisms in Delivery \cite{ipc2023numeric}.
Safety verification determines \emph{whether an unsafe marking is reachable from an initial marking}.
If the unsafe markings are unreachable, the system is safe.
The tested problems vary in size.
Model checkers have a $10$ minute timeout and $8$-GiB memory limit.

TMKit's relaxation outperforms ITS-tools in safety verification on trace nets (see \autoref{img:inftime}) and performs comparably to ITS-tools on ordinary PNs (see \autoref{table:mcc-false-ef}).
Our approach proves 38 unreachabilities out of 40 instances with trace nets, while ITS-tools solves 25 instances.
In addition, TMKit scales better than ITS-tools as we increase the problem size.
Finally, TMKit verifies approximately the same number of unreachabilities as ITS-tools in instances with ordinary PNs.

The difference in performance between TMKit and ITS-tools is due to TMKit's sound and incomplete relaxation and ITS-Tools' combination of state-space computation and incomplete subroutines for proving unreachabilities.
TMKit outperforms ITS-tools in the tested problems with trace nets as
our relaxation effectively proves unreachabilities through efficiently-solved linear programs, whereas ITS-tools' incomplete techniques do not support trace nets, so ITS-tools fully rely on the expensive state-space computation.
In contrast, ITS-tools' complete procedure and its support techniques guarantee ITS-tools to find more unreachabilities in ordinary PNs than TMKit's incomplete relaxation.

\subsection{Invariant Generation} \label{exp:inv}
\begin{table}[t]
\centering
\scriptsize
\setlength{\tabcolsep}{3pt}
\begin{tabular}{@{}c|l|rr|rr|rr|rr@{}}
\hline
& & \multicolumn{2}{c|}{\cite{sting2004}}
& \multicolumn{2}{c|}{\cite{sankaranarayanan2003petri}}
& \multicolumn{2}{c|}{Polynomial}
& \multicolumn{2}{c}{MILP} \\
\cline{3-4} \cline{5-6} \cline{7-8} \cline{9-10}
& & Count & Time & Count & Time & Count & Time & Count & Time \\
\hline
\multirow{5}{*}{\textbf{1C}}
& VisitAll (18)    &  \best{1} & 5.93 & \best{1} & 0.04 & \best{1} & 0.03 & \best{1} & 0.93 \\
& Blocksworld (19) &  0 & 13.17 & 7 & 0.06 & 13 & 0.02 & \best{19} & 0.22 \\
% & Satellite (47)   & -- & --   & \best{2} & 0.12 & \best{2} & 0.07 & \best{2} & 0.09 \\
& Delivery (49)    & -- & --   & \best{10} & 1.11 & \best{10} & 0.33 & \best{10} & 0.68 \\
& Rover (324)      & -- & --   & \best{4} & 0.15 & \best{4} & 0.05 & \best{4} & 0.08 \\
\hline
\multirow{5}{*}{\textbf{1H}}
& VisitAll (18)    &  \best{1} &  5.93 & \best{1} & 0.05 &  0 & 0.03 & \best{1} & 0.95 \\
& Blocksworld (19) &  0 & 13.17 &  0 & 0.05 & \best{7} & 0.02 & \best{7} & 0.11 \\
% & Satellite (47)   & -- & --   &  0 & 0.13 & \best{2} & 0.07 & \best{2} & 0.09 \\
& Delivery (49)    & -- & --   &  0 & 0.90 & \best{10} & 0.33 & \best{10} & 0.73 \\
& Rover (324)      & -- & --   &  0 & 0.15 & \best{4} & 0.05 & \best{4} & 0.08 \\
\hline
\end{tabular}
\caption{Invariant counts and generation times among the baselines \cite{sting2004,sankaranarayanan2003petri}, our polynomial-time algorithm, and our MILP extraction for $1$-cardinality (denoted $\mathbf{1C}$) and $1$-hot (denoted $\mathbf{1H}$) invariants. The number in each problem shows the place count. Dashes represent 10-minute timeout.}
\label{table:invariant}
\end{table}

We evaluate our polynomial-time invariant generation in \autoref{inv-mutex} and our MILP invariant generation in \autoref{inv-milp} on four different trace nets, modeling four different problems from the past IPCs \cite{ipc2000,ipc2014,ipc2023numeric}.
% \cite{ipc2000,ipc2014,ipc2004,ipc2023numeric}.
We generate $1$-cardinality and $1$-hot invariants in the tested problems.
We report nontrivial invariants (excluding, e.g., box constraints on Boolean variables) and count $1$-hot invariants as $1$-cardinality invariants in \autoref{table:invariant}.
We compare our approaches against StInG \cite{sting2004} on the tested trace nets and against \cite{sankaranarayanan2003petri} on modified versions of the same nets because \cite{sankaranarayanan2003petri} applies only to ordinary PNs.

We transform the tested trace nets to ordinary PNs to apply \cite{sankaranarayanan2003petri}.
For \cite{sankaranarayanan2003petri} to produce correct invariants in the original trace nets, the set of reachable markings in the resulting net must be a super-set of the original reachable markings.
We ensure the condition by adjusting the two TN flow relations ($\FuncSty{set}$ and $\FuncSty{reset}$), which are not in ordinary PNs, with dummy transitions.
For each Boolean place in a $\FuncSty{set}$ ($\FuncSty{reset}$) relation, we create an outgoing (incoming) transition from (to) the place with a $1$-weighted arc and let the Boolean place be an integer place.
The place accumulates (removes) tokens from transitions in $\FuncSty{set}$ ($\FuncSty{reset}$) relations, while the dummy transition restores its token count to one (zero), as may be required for other neighboring transitions to fire.
Dummy transitions ensure the resulting net preserve the original firing sequences and, therefore, the original reachable markings.

Our invariant generations outperform both baselines in the number of invariants found and the generation times (see \autoref{table:invariant}).
Our polynomial-time algorithm in \autoref{inv-mutex} matches or exceeds the baselines on $1$-cardinality invariants and, in most cases, more $1$-hot invariants, and has the lowest runtime on all benchmarks.
Compared to all methods, our MILP approach in \autoref{inv-milp} produces the most invariants---which subsume others---while remains slightly slower than our polynomial-time algorithm.

TMKit's relaxation-based methods and the baselines' induction methods, together with their different extraction techniques, lead to different performance.
Our polynomial-time algorithm has the best runtime as it uses constraint solvers and a greedy technique to efficiently grow linear invariants from our relaxation.
Our MILP approach produces the most invariants as traps tighten the relaxation, at the expense of longer runtime relative to our polynomial-time algorithm.
StInG \cite{sting2004} produces the least invariants due to accuracy-loss as a tradeoff for resolving non-linearity in its formulated constraints for solving \cite{sankaranarayanan2004constraint} and scales poorly due to the expensive constraint processing and solution extraction \cite{sankaranarayanan2004constraint,colon2003linear}.
While \cite{sankaranarayanan2003petri} resolves the non-linearity in \cite{sting2004} without accuracy-loss, its constraint formulation for solving is rigid and is unable to account for known useful information, e.g., box constraints on Boolean places and traps, resulting in fewer invariants found compared to our approaches.
%

% All four approaches construct a set of constraints whose solutions correspond to invariants; however, we derive the constraints from our relaxation, whereas \cite{sting2004,sankaranarayanan2003petri} use induction, asserting an invariant template that holds before and after every transition.
% Linear relaxation is more effective in finding invariants than the induction approach in the tested problems for trace nets.

\section{Conclusions}
We propose a reachability relaxation and two methods for finding affine invariants in hybrid trace nets.
% : a polynomial-time algorithm and a comprehensive enumeration with MILP.
%
Our reachability relaxation is a linear program; reachability in LP is necessary for reachability in trace nets, offering polynomial time invariant generation and fast unreachability verification.
Strengthening the relaxation with traps allows identification of invariants missed by the polynomial-time invariant generation.
%
% Farkas' lemma permits extraction of affine invariants held in our strengthened relaxation.
%
MILP solvers provide efficient tools for incrementally solving $k$-cardinality and $k$-hot invariants in our strengthened relaxation.
Empirically, our linear relaxation proves $1.5\times$ more unreachabilities and scales better than the baseline in the tested problems.
Our invariant generations produce $2.5\times$ more invariants and run up to $2$ order of magnitude faster than the baselines in the tested problems.

% 1.5x more safety
% 2.5x more invariants
% up to 2 order of magnitude faster
% Invariants are useful for reducing verification time through state-space search or construction.
%
\bibliographystyle{plain}
\bibliography{dyalab-refs,dyalab-pubs,cit}

\end{document}

%% file: tables/boolean-table.tex
%\renewcommand{\arraystretch}{1.4}
%\newcolumntype{C}[1]{>{\centering\arraybackslash}p{#1}}
\setlength{\extrarowheight}{0pt}
\begin{table}[t]
\centering
\scriptsize
%\begin{tabular}{|C{1cm} | C{1cm} | C{1cm} | C{1cm} | C{1cm}|}
\begin{tabular}{rccrr}
%\hline
  %\rule{0pt}{1.1em}
  Relations & $\markingstepvar{k}{i}$ &
                                                      $\markingstepvar{k+1}{i}$ & $\wfunc(\placevar{i}, \trans{j})$ & $\wfunc(\trans{j}, \placevar{i})$
  \\
\hline
\FuncSty{input} & $1$ & $0$ & $1$ & $-$  \\
%\hline
\FuncSty{output} & $0$ & $1$ & $-1$ & $-$  \\
%\hline
\FuncSty{used} & $1$ & $1$ & $1$ & $1$ \\
%\hline
\FuncSty{free} & $0$ & $0$  & $-1$ & $-1$ \\
%\hline
\FuncSty{set} & $0 \lor 1$ & $1$ & $-$ & $1$  \\
%\hline
\FuncSty{reset} & $0 \lor 1$ & $0$ & $-$ & $-1$   \\
%\hline
\FuncSty{nop} & $0 \lor 1$ & $0 \lor 1$ & $-$ & $-$  \\
%\hline
\end{tabular}
\caption{Trace Net flow relation semantics and weights.}
\label{table:boolean}
\end{table}

%%% Local Variables:
%%% mode: latex
%%% TeX-master: "../paper"
%%% End: